\documentclass[12pt]{article}
\usepackage{graphicx} % for easy postscript figure inclusion.
\usepackage{rotating}
\usepackage{epstopdf}

\begin{document}

% these are abbreviations for things we might use many times later:
 
\title{Expert Witness Report: \\ Regina v. Hubbs \\ Determination of the Sun's Position}
\author{ Raminder Singh Samra\\
\\
\date{July 23, 2010}
\\
HR MacMillan Space Centre\\
1100 Chestnut Street\\
Vancouver B.C\\
Canada, V6J 3J9\\
\\
%Department of Physics and Astronomy, \\University of British Columbia \\
%             6224 Agricultural Road, \\Vancouver, British Columbia, \\Canada, V6T 1Z1
}
\maketitle
\pagebreak
\begin{abstract}
The Sun's position in the equitorial coordinate system was calculated for February 9th 2009  at 13:00 PST.  The coordinates were then converted to local coordinates for the observer located at $49.41^{\circ}$ N and -$123.58^{\circ}$ W. The Sun's position was found to be Alt: $\sim 26^{\circ}$, Az: $\sim189^{\circ} $. This position was found to not impede the vision of a driver headed east on the Sunshine Coast Highway at the intersection of Conrad Road. 

\end{abstract}
%\pagebreak

\section{Introduction}

The position of the Sun can be expressed in a coordinate system commonly referred to as the equatorial coordinate system. The system allows observers to describe locations of objects in they sky using a pair of numbers called right ascension and declination. A given star has roughly the same equatorial coordinates in contrast to the local horizontal coordinates which are based upon geographical latitude and longitude of the observer; and constantly change based upon time of day.  The equatorial coordinates of objects such as stars and distant deep sky objects vary over long periods of time, however for closer objects such as the sun, moon and planets the equatorial coordinates vary on a daily basis. It is possible to calculate the equatorial coordinates of any given planet or the Sun by using advanced geometry and knowing a handful of orbital parameters.  These parameters are known as the ephemeris and can be found in various sources and astronomical almanacs.

\begin{figure}[h]
\centering
\includegraphics[angle=0,width=0.75\textwidth]{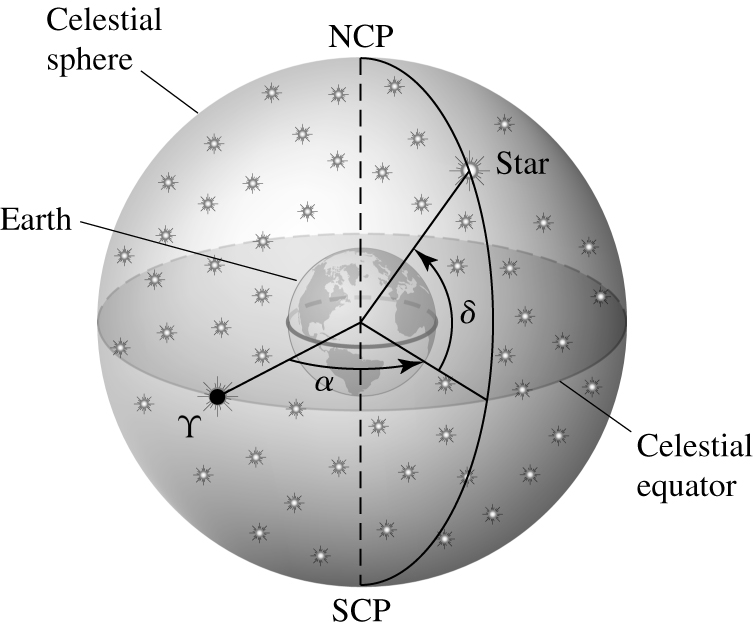}
\caption{The equatorial coordinate system. $\alpha$, $\delta $, $\Upsilon $ designate right ascension, declination and the position of the vernal equinox, respectively. NCP and SCP are the north and south celestial poles. Figure from Carroll and Ostlie \cite{co}}
\label{fig:eq}
\end{figure}

\section{Determining the Sun's Position}

The position of the Sun as seen from the Earth is determined by four quantities.  The first is the time; the Sun's position in the equatorial coordinate system varies on a daily basis so therefore it is important to know the precise time to calculate its position. The second quantity is the motion of the Earth in its orbit around the Sun, which does not happen at a constant speed due to the eccentricity of the Earth's orbit. The third quantity is the angle between the axis of rotation of the Earth and the plane of the Earth's orbit.  This is close to $23^{\circ} $ which is responsible for the seasons. Finally, the observer's location on the Earth which determines how high in the sky the Sun can get.

\subsection{Calculations}

We will now calculate the Sun's position at 13:00 (1:00pm) on February 9th 2009.  As we see the Sun from a planet, the motion of the Earth around the Sun is reflected in the apparent motion of the Sun along the ecliptic plane relative to the stars. If the orbit of the Earth were a perfect circle then the Sun as seen from the Earth would move at a fixed speed and it would be simple to calculate its position. The position that the Earth has relative to its perihelion (closest approach to the Sun) if the orbit was circular is called the mean anomaly and it is approximated in Eqn - \ref{eqn:mean}.  In Eqn -\ref{eqn:mean} $ d $ is measured in whole and fractional days since 00:00 GMT on the most recent January 1st.

\begin{equation} \label{eqn:mean}
M = -3.59^{\circ} + 0.98560^{\circ} d 
\end{equation}

For the chosen date d = 40.875, $M = 36.6964^{\circ} $

The orbits of the planets are not perfect circles but rather ellipses so the speed of the planet varies and so does the Sun's speed across the sky.  The true anomaly is represented by Eqn - \ref{eqn:ano}, here $\nu $ is the angular distance between the Earth from its perihelion as seen from the Sun. For a circular orbit the mean and true anomaly are the same. Eqn - \ref{eqn:C} is the difference between the mean and true anomaly and is called the Equation of Centre.

\begin{equation} \label{eqn:ano}
\nu = M + C
\end{equation} 

Where 

\begin{equation} \label{eqn:C}
C \approx C_{1} Sin(M) + C_{2} Sin(2 M) + C_{3} Sin(3 M) + C_{4}Sin(4M) + C_{5}Sin(5M) 
\end{equation} 

For the Earth the values of $C_{1} $ thru $C_{3} $ are: 1.9148, 0.0200, 0.0003 respectively and $C_{4} $ and $C_{5} $ being negligible.  Using the value of $M $  obtained from Eqn - \ref{eqn:mean} we get a value of $\nu  = 37.8601^{\circ} $. 

To find the position of the Sun in they sky we need to know what the ecliptic longitude $\Pi $ is of the perihelion of the Earth relative to the ecliptic and vernal equinox of the Earth. The ecliptic of the Earth is the plane of the orbit and the vernal equinox is the point where the Sun passes from south to north through the plane of the Earth's equator. Additionally we need to know the obliquity $\epsilon $ of the Earth's equator compared to the orbit. These two values are measured in degrees and are $\Pi $ = 102.9372, $\epsilon $ = 23.45. 

The ecliptic longitude $\lambda $ is the position along the ecliptic relative to the vernal equinox, this as seen from the Sun is given by Eqn - \ref{eqn:ecp}. 

\begin{equation} \label{eqn:ecp}
\lambda = \nu + \Pi = M + \Pi + C
\end{equation}

If you look at the Sun from the Earth then you are looking exactly the opposite direction if you look at the Earth from the Sun, so that difference is $180^{\circ} $. The ecliptic longitude of the Sun as seen from the Earth is equal to:

\begin{equation} \label{eqn:sun}
\lambda_{sun} = \nu + \Pi + 180^{\circ} = M + \Pi + C + 180^{\circ} 
\end{equation}

\begin{figure}[h]
\centering
\includegraphics[angle=0,width=1\textwidth]{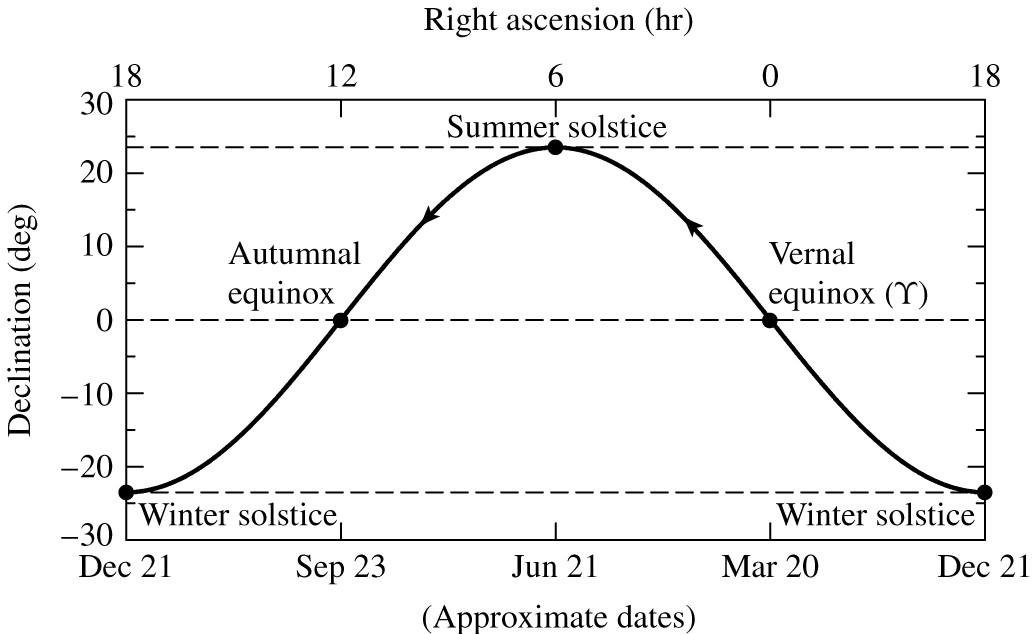}
\caption{The ecliptic is the annual path of the Sun across the celestial sphere and is a sinusodial curve about the celestial equator. See Fig - \ref{fig:eq} for a view of the celestial sphere. Figure from Carroll and Ostlie \cite{co}}
\label{fig:epl}
\end{figure}

The value of $\lambda_{sun} $ determines when the seasons begin. When $\lambda_{sun} $ = $0^{\circ}$ that is the beginning of the spring in the northern hemisphere.  The value of $\lambda_{sun} $ from Eqn - \ref{eqn:sun} was found to be $ 320.7973^{\circ} $. This value represents a point on the curve in Fig - \ref{fig:epl}; from this it is possible to extract the right ascension and declination of the Sun.

\subsection{The Equatorial Coordinates}

The equatorial coordinate system is tied to the rotation axis of the Earth; the coordinates are right ascension $\alpha $ (alpha) and declination $\delta $ (delta). The declination determines from which parts of the Earth the Sun is visible and the right ascension determines primarily when the object is visible.  
For the Sun we have: 

\begin{equation} \label{eqn:ra}
\alpha_{sun} = tan^{-1}[ \frac{sin( \lambda_{sun} )\cdot cos (\epsilon) } { cos ( \lambda_{sun}) }]
\end{equation}

\begin{equation}\label{eqn:dec}
\delta_{sun} = sin^{-1} [ sin(\lambda_{sun}) \cdot sin(\epsilon)]
\end{equation}

Using Eqns - \ref{eqn:ra} and \ref{eqn:dec} we get the Sun's coordinates to be $\alpha $ =  21h:32m:46s  and $\delta $ = -$14^{\circ}32.09 $ for 13:00 PST on February 9th 2009. Additionally using the same procedure for 12:45 PST on February 9th 2010 we get $\alpha $ = 21h:33m:17s and $\delta $ = -$14^{\circ}29.97 $.  A difference of 31 seconds in right accession and 2.12 minutes of arc in declination. 

\begin{figure}[h]
\centering
\includegraphics[angle=0,width=0.7\textwidth]{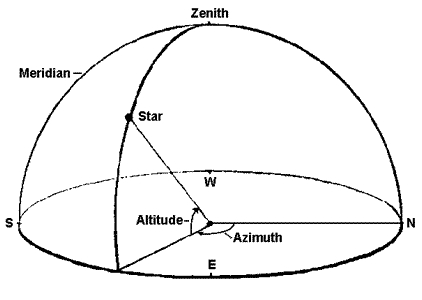}
\caption{The local horizon coordinate system. On the date and time in question the Sun was found to be at an altitude of $25.7^{\circ} $ and an azimuth angle of $189.4^{\circ} $}
\label{fig:altAzz}
\end{figure}

Finally we convert the right accession and declination into the local coordinate system (Fig - \ref{fig:altAzz}) to calculate the position of the Sun relative to an observer at $49.41^{\circ}$ N and -$123.58^{\circ}$ W corresponding to the coordinates at the intersection of the Sunshine Coast Highway and Conrad Road.  We first get the hour angle $H $, that is the time since the Sun passed through the celestial meridian.  H = Local Sideral Time - Right Ascession; the sidereal time at the time in question was 22h:08m. The conversion from equatorial to horizon coordinates are given below:

\begin{equation} \label{eqn:alt}
sin(a) = sin(\delta)sin(\phi) + cos(\delta)cos(\phi)cos(H)
\end{equation}

\begin{equation} \label{eqn:az}
cos (A) = \frac{ sin(\delta) - sin(\phi)sin(a)}{cos(\phi)cos(a)}
\end{equation}

Here $ a $ is the altitude of the Sun, $\phi $ is the observer's latitude, $H $ is the hour angle, $ A $ is the azimuth angle of the observer measured eastward from north.  Using Eqns - \ref{eqn:alt} and \ref{eqn:az}  and converting to the conventional coordinate system and quadrants we get an altitude of $ 25.7^{\circ} $ and azimuth of 189.4 degrees east of north, corresponding to 9 degrees west of due south.

\begin{figure}[h]
\centering
\includegraphics[angle=0,width=0.8\textwidth]{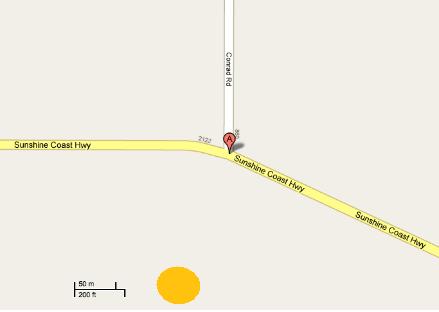}
\caption{The apparent position of the Sun is drawn as the yellow circle for an observer at the intersection of Conrad Road and Sunshine Coast Highway. Note that this drawn position is only valid for an observer at the intersection and due to parallax one cannot assume that the Sun's position changes across the map; the Sun will always be at the same local horizon coordinates for an observer across this map.  North is directly parallel to Conrad Road. The Sun here is about nine degrees west of south. Image from Google Earth.}
\label{fig:map}
\end{figure}

\begin{figure}
\centering
\includegraphics[angle=0,width=0.65\textwidth]{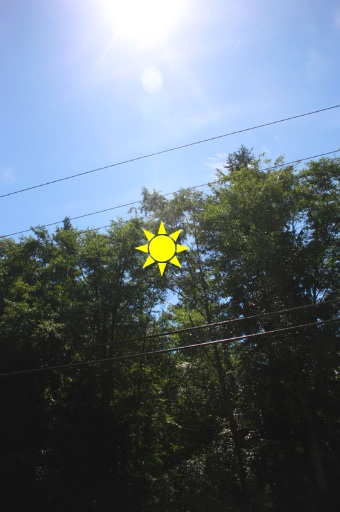}
\caption{The apparent position of the Sun is drawn in here for February 9th 2010 at 1pm. The image shows the correct azimuth of the Sun but the altitude is different.  This is due to the images captured months apart. The Sun's position is slightly above the tree line, therefore it is reasonable to assume if the trees were evergreens, the rays would not penetrate through the tree line; shading the road. However as some of the trees are not evergreens it is not a valid assumption that the entire road was in shade at the date in question.}
\label{fig:tree}
\end{figure}
\section{Discussion}

\begin{sidewaysfigure}
\centering
\includegraphics[angle=0,width=1.1 \textwidth]{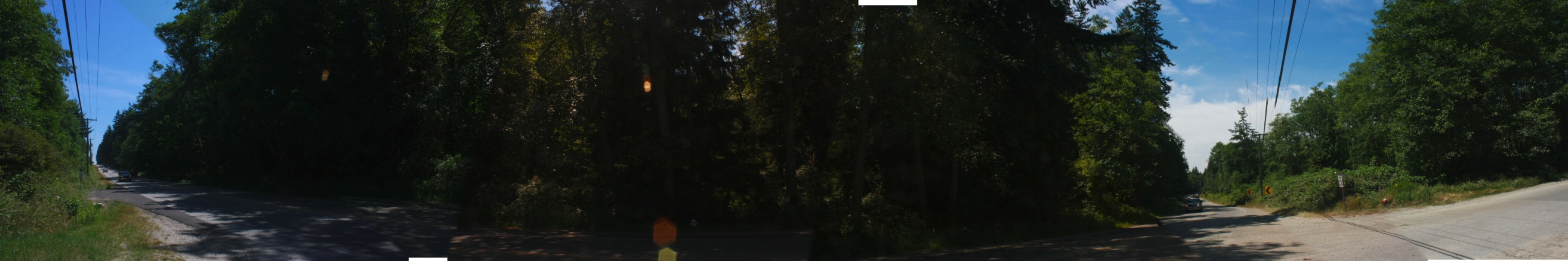}
\caption{A panorama of the site showing the bend in the road and the amount of shade the trees provide when the Sun's altitude is $60^{\circ}$ and azimuth is $\sim189^{\circ} $. On the date in question the Sun was far lower and below the tree-line. Regardless of the altitude of the Sun at this azimuth it is apparent that the Sun would not impair the vision of drivers headed eastbound. In this image eastbound towards Gibsons is on the left, Conrad Road can be seen on the far right.  }
\label{fig:pano}
\end{sidewaysfigure}

We now focus on the Sun's position in the sky at 13:00 PST February 9th 2009 at the intersection of Conrad Road and the Sunshine Coast Highway and how it could possibly impair the vision of any motorists. Figure - \ref{fig:map} shows the apparent position of the Sun from the intersection of Conrad Road and the Sunshine Coast Highway.

Additionally the position of the Sun one year later on February 9th 2010 at 12:45 PST was found to be $25.2^{\circ} $ and an azimuth angle of $185.4^{\circ} $. This difference is very minimal and is due to the fifteen minutes between observations and is difficult to distinguish for any observer at the site. Therefore any photographs of the sky obtained in terms of the Sun's position are accurate  on the same date and time one year later.

For a motorist traveling eastbound towards Gibsons on the Sunshine Coast Highway the Sun would be on the right side visible from only the passenger side window (Figs - \ref{fig:tree} , \ref{fig:pano}).  From this it is evident that the Sun would not be in the central or peripheral vision of a driver. The only way the Sun could impair the vision of any motorist was if one was headed south on Conrad Road, even in that situation the motorist could use the visors in the car to prevent the rays from distracting the driver.

\section{Conclusions}

The position of the Sun was calculated for February 9th 2009 at 13:00 PST and was found to be at Alt: $\sim 26^{\circ} $ and  Az: $\sim 189^{\circ} $. This position of the Sun was found not to impair the vision of any driver headed eastbound on the Sunshine Coast Highway. Additionally the Sun's position was calculated one year later and it was found to be similar to the position one year earlier.

\end{document}